\documentclass[10pt, twocolumn]{article}

\newcommand{\ket}[1]{\left| #1\right\rangle}

\newcommand{\eqref}[1]{(\ref{#1})}

\usepackage{ol2}

\begin{document}

\twocolumn[

\title{Matched Slow Pulses Using Double Electromagnetically Induced Transparency}

\author{Andrew MacRae, Geoff Campbell, A. I. Lvovsky}
\address{Institute for Quantum Information Sciences, University of Calgary, Alberta T2N 1N4, Canada}

\date{\today}

\begin{abstract}
We implement double electromagnetically-induced transparency (DEIT) in rubidium vapor, using a tripod-shaped energy level scheme consisting of hyperfine and magnetic sublevels of the $5S_{1/2} \rightarrow 5P_{1/2}$ transition. We show experimentally that through the use of DEIT one can control the contrast of transparency windows as well as group velocities of the two signal fields. In particular, the group velocities can be equalized, which holds promise to greatly enhance nonlinear optical interaction between these fields.
\end{abstract}

\maketitle
]

Electromagnetically induced transparency (EIT) \cite{EIT_review} has become a cornerstone of many methods for controlling optical fields. The ability to slow or even store pulses of light has applications ranging from precision interferometry to all-optical buffers in classical and quantum information networks. An extension of EIT, double EIT (DEIT), has been proposed and demonstrated as a vehicle for extending the utility of EIT schemes by creating transparency conditions for two signal fields simultaneously \cite{GNL, Cap_DEIT, Rebic_99,Petrosyan_04}. This provides the possibility for coherent control \cite{Lukin_01, Scully, RATOS} and nonlinear interaction between weak optical fields \cite{GNL, Cap_DEIT}. Double EIT allows propagation of the two signal fields with minimal loss, and increases the interaction time between pulses due to group velocity reduction. These capabilities make DEIT a promising candidate for numerous applications in quantum computation and communication.

In this paper we demonstrate a DEIT system and investigate the interdependency of the two signal fields in both continuous-wave (CW) and pulsed cases. In the CW case the effect of optical pumping due to the second signal field is investigated. We observe that it can be used to enhance the contrast of the transparency window. In the pulsed regime, we demonstrate that signal fields can be slowed and stored with little impact on each other.

One of the experimental challenges in the application of DEIT for achieving optical nonlinearities is
that the group velocities of the two signal pulses are, generally, not equal. This may reduce their interaction time and reduce the nonlinearity. In this work, we show that the delay induced in the pulses can be adjusted and matched through preparation of the atomic states. In addition to quantum computational gates, this technique could find application in quantum communication protocols that use frequency multiplexed information channels, such as simultaneous quantum memory, simple qubit operations and correction of time delays \cite{Khurgin}.

\begin{figure}
 \centering
 \includegraphics[]{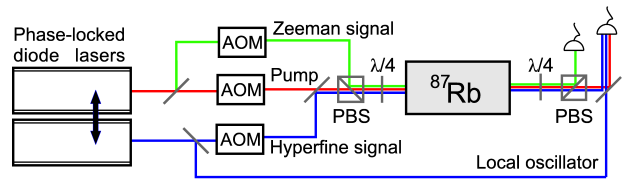}
\caption{(Color online) Simplified experimental setup.}
\label{ExpSet}
\end{figure}

\begin{figure}
\centering
 \includegraphics[]{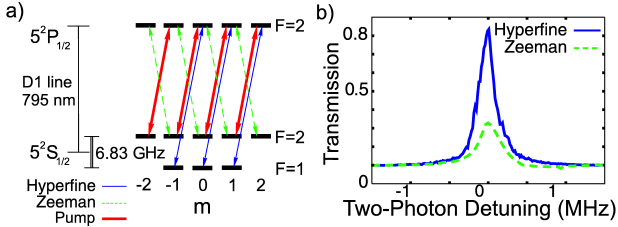}
\caption{(Color online) Left: the atomic level scheme used in this experiment is a tripod scheme with both hyperfine levels of the $5^2S_{1/2}$ shell as ground states and the $F=2$ hyperfine state was of of the $5^2P_{1/2}$ shell as the excited state. Right: simultaneous EIT windows observed by scanning the pump field.}
 \label{Levels}
\end{figure}

The experimental setup is displayed in Fig.~\ref{ExpSet}. The experiments were performed in a magnetically shielded $12$-cm vapor cell containing $^{87}$Rb gas and $10$ torr Ne buffer gas maintained at $45^\circ$C. The atomic level system was of tripod shape, formed by both hyperfine levels of the $5^2S_{1/2}$ shell as ground states and the $F=2$ hyperfine state of the $5^2P_{1/2}$ shell as the excited state (Fig.~\ref{Levels}(a)). The pump field was $\sigma^+$ polarized and coupled the $F=2$ ground level to the $F'=2$ excited level. The \textit{Zeeman} signal field coupled the same levels as the pump, but was of $\sigma^-$ polarization. The second signal field --- the \textit{hyperfine} field, was $\sigma^+$ polarized and coupled the $F=1$ ground level to the $F'=2$ excited level. The hyperfine splitting of the excited levels is $817$ MHz which is much more than the 275 MHz Doppler-broadened width of the optical transition. Hence the $F=1$ hyperfine excited level did not affect the experiment.

The 795 nm laser fields were produced by two self-made external cavity diode lasers. In order that the two lasers acted coherently on a single set of atoms, they were phase locked so that their frequency difference remained fixed at $6.834$ GHz \cite{PLL}. One of the lasers provided the pump and Zeeman fields; the other laser generated the hyperfine field and the local oscillator for its detection. The pump and each of the signal fields passed through acousto-optic modulators (AOMs) allowing them to be scanned in frequency, or switched on or off independently, which permits the creation of pulses. The beams were then spatially mode matched and passed through a quarter-wave plate to provide circular polarization. Typical laser beam power in this experiment amounted to 2.5 mW for the pump, and 1--150 $\mu$W for the signal fields. The beam diameters in the cell were about 750 $\mu$m.

After passage through the atomic medium, another quarter-wave plate and a polarizing beam splitter (PBS) separated the fields into two paths; one with the Zeeman field, and the other containing the hyperfine and pump fields.
The Zeeman field was measured directly with a photodiode. Since the weak hyperfine field was in the same spatial and polarization mode as the strong pump, it could not be spatially separated. To measure this field, we employed heterodyne detection with a local oscillator differing in frequency by 200 MHz. The resultant beat frequency was observed with a spectrum analyzer in a time-resolved setting with a resolution of 200 ns.

In order to observe the EIT dip simultaneously for both signal fields in the CW regime, the pump field frequency was scanned, effectively varying over two-photon detuning for each signal field. A plot of double EIT is shown in Fig.~\ref{Levels}(b).

\begin{figure}
 \centering
 \includegraphics[]{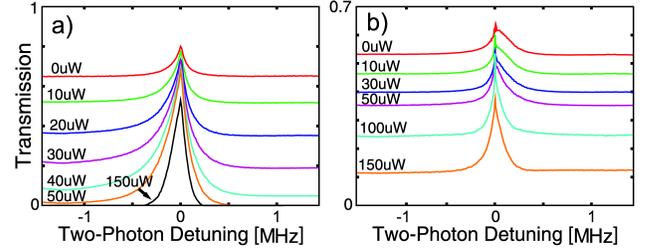}
\caption{(Color online) The effect of one signal field on the other in the DEIT system. In (a), the absorption spectrum of the hyperfine field, fixed at 50 $\mu$W, is measured for various powers of the Zeeman EIT signal. In (b), the Zeeman signal is of constant intensity and the hyperfine power is varied.}
 \label{DEIT}
\end{figure}

In the CW case, there was a significant effect of one signal field on the other, even for moderate field strengths. Specifically, when the power of one signal field was increased, the transparency contrast in the other enhanced (Fig.~\ref{DEIT}). This can be explained by optical pumping from one ground state to the other. From the atomic level scheme (Fig.~\ref{Levels}), we see that when the hyperfine field is absent, the atomic population will collect in the $F=1$ ground state. Turning on the hyperfine field will now serve to repopulate the $F=2$ ground state, specifically the $m=2$ sublevel, increasing the effective atomic density of atoms experiencing EIT. Similarly, with only the hyperfine field present, the $|F=2,m=2\rangle$ ground state will accumulate, and turning on the Zeeman field will have the effect of increasing the number of atoms in the $F=1$ ground state. Although the overall contrast of the EIT window improves in the presence of the repumping field, the transparency at the two-photon resonance also somewhat reduces \cite{EIT_decoh}.

Next, the slow light effect was observed on each signal field. The signal field AOMs were operated to produce simultaneous 1 $\mu$s pulses. Although each field experienced significant group delays, the group velocities were not equal because of differences in ground state populations and transition dipole moments. However, by initially pumping the atoms from one ground state to another, we were able to change the population of each state, thus reducing the group velocity of one field while increasing that of the other. The amount of population transfer depends on both duration and the power of the preparation pumping. By selecting appropriate values for each, the group velocities of the pulses could be matched.

\begin{figure}
 \centering
 \includegraphics[]{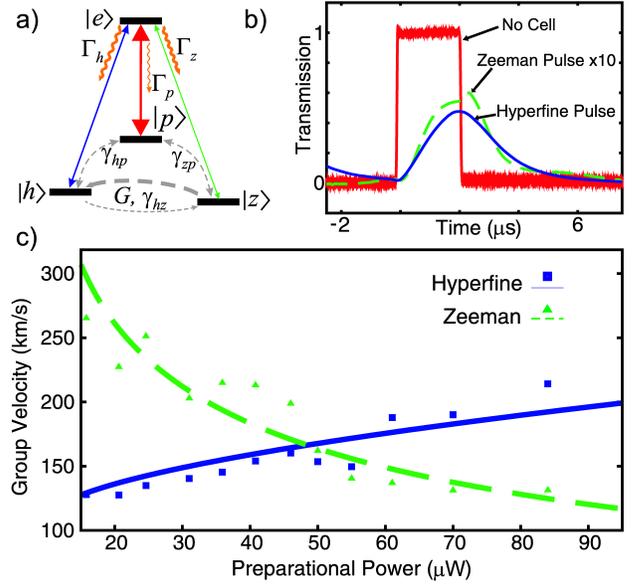}
\caption{(Color online) Group velocity matching. (a) The theoretical tripod scheme analyzed in the text. (b) Pulse waveforms with matched group velocities. The tail of the preparation hyperfine pulse (55 $\mu$W) is visible. (c) Group velocities of the co-propagating signal pulses as a function of the preparation power. Solid lines: theoretical model (see text).}
\label{Group_Vel_Curve}
\end{figure}

The initial preparation was performed by switching on one of the signal fields for some time prior to sending the pulses. Using a 500 $\mu$s preparation pulse in the hyperfine field and varying its power, we found that the group velocities are matched at $135$ km/s in the neighborhood of 50 $\mu$W preparation pulse power (Fig.~\ref{Group_Vel_Curve}). Conversely, the preparation pulse can be in the Zeeman field, having the effect of decreasing the group velocity of the hyperfine pulse. By using a strong (150 $\mu$W, 500 $\mu$s) preparation pulse, we were able to enhance the delay by an additional factor of $10$, achieving a delay of several pulse lengths.

In order to gain understanding of our observations, we analyzed a model of a tripod atomic system consisting of a common excited state $\vert e \rangle$ and three ground states $\vert z \rangle$, $\vert h \rangle$, $\vert p \rangle$, coupled by two weak signal fields $\Omega_{z}$, $\Omega_{h}$, and a pump $\Omega_{p}$ respectively. The decoherence mechanisms included dephasing (decay of the off-diagonal matrix elements) $\gamma_{ij}$ acting within each pair of ground states, and population exchange $G$ between levels $\ket h$ and $\ket z$ (Fig. \ref{Group_Vel_Curve}(a)). Applying the above decoherence parameters to the light-atom Hamiltonian in the rotating wave approximation, the density matrix in the presence of the pump and preparation (one of the two signal) fields was obtained by solving the Liouville equation in the steady state. The group velocities of the signal pulses have then been calculated using the linear response theory with the initial state given by the steady state density matrix. This approach is justified because the period when the preparation pulse was off (10 $\mu$s) was much shorter than the relaxation time of the atomic system (measured to be on the order of several hundred microseconds).

The susceptibility of the atomic gas for the two signal fields was evaluated numerically for varying strengths of the preparation field. The parameters of the system were set to fit the group velocity behavior to that observed experimentally. A good fit was obtained with a Doppler width of $500$ MHz, $\Gamma_{z}=\Gamma_{h}=\Gamma_{p}=6$ MHz, $\gamma_{hz}=\gamma_{hp}\sim 5$ kHz. $\gamma_{zp}\sim 40$ kHz, and $G = 50$ Hz. In the CW picture, our theoretical model has also been able to qualitatively reproduce the enhancement of the transparency contrast for one signal field when increasing the strength of the other (Fig.~\ref{DEIT}).

We found that a crucial parameter in obtaining a good fit was the decoherence between the two signal ground states $\ket h$ and $\ket z$. As $\gamma_{hz}$ was increased, the optical pumping effect became much more significant. With small $\gamma_{hz}$, enhanced transparency on 2-photon resonance in the CW regime emerged due to the presence of an additional dark state formed by energy levels $\ket h$ and $\ket z$. In the experiment, a small transparency dip corresponding to this dark state was indeed observed when the two signal fields were at a two-photon resonance with each other.

Also interesting is the role of the population exchange rate $G$. As was found previously \cite{EIT_decoh}, the fraction of this mechanism in ground state decoherence is small compared to pure dephasing $\gamma_{ij}$. This finding is confirmed by the present experiment. Yet the population exchange mechanism cannot be completely neglected. If we set $G=0$, the populations of states $\ket h$ and $\ket z$ would depend only on the ratio between $\Omega_h$ and $\Omega_z$, but not on their absolute magnitude. The nonlinear effect of the signal fields on each other could then be observed at arbitrarily low field strength, which is not in agreement with our experimental observation. This discrepancy can be addressed by setting a small, but nonzero value of $G$, which governs the ground state populations at low signal intensities. 

Intuitively the dynamics of the system can be understood by expressing the Zeeman and hyperfine ground states as linear combinations of ``dark'' and ``bright'' states: $\ket d = \Omega_h \ket z - \Omega_z \ket h$ and $\ket b = \Omega_z \ket z + \Omega_h \ket h$, respectively. Of these only the bright state couples to $\ket e$. Hence, in the absence of population exchange between $\ket b$ and $\ket d$, the linear susceptibility is determined only by the composition of the bright state, i.e. the ratio between $\Omega_z$ and $\Omega_h$.

\begin{figure}
 \centering
 \includegraphics[]{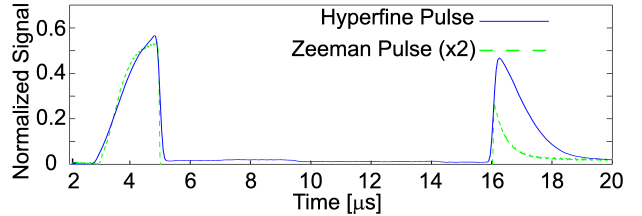}
\caption{(Color online) Simultaneous storage and retrieval of two signal pulses using a single pump field.}
\label{Storage}
\end{figure}

By switching off the pump field during the propagation of the pulses, we were able to simultaneously store two pulses and retrieve them at a later time by switching the pump field back on. Storage times of more than 100 $\mu$s were observed, with decreasing pulse retrieval efficiency for longer times.  Figure \ref{Storage} displays a simultaneous storage run with a storage time of 10 $\mu$s.

In summary, we have realized simultaneous group velocity reduction and storage for two signal fields  using double EIT.  Furthermore, we have demonstrated that the relative group velocities of the signal pulses can be varied and even matched. The observations are confirmed by a simple theoretical model. This development has potential applications in coherent control of multiple fields and increased interaction time for pulsed non-linear schemes. The observed simultaneous storage of two modes may find application in 
classical optical switching.

We gratefully acknowledge K.-P.Marzlin for fruitful discussions.
This work was supported by NSERC, CIAR, iCORE, AIF, CFI and Quantum\emph{Works}.

\bibliographystyle{unsrt}

\end{document}